\documentclass{emulateapj}
\usepackage{graphicx}
\usepackage{natbib}

\shorttitle{Eri~II and its star cluster}
\shortauthors{Crnojevi\'c et al.}

\newcommand{\kms}{\ensuremath{\rm{km\,s^{-1}}}}
\newcommand{\Msun}{\ensuremath{M_\odot}}
\newcommand{\MLsun}{\ensuremath{M_\odot/L_\odot}}
\newcommand{\Mlim}{\ensuremath{M_{HI}^{lim}}}
\newcommand{\gasfrac}{\ensuremath{{M_{HI}/L_{V}}}}

\begin{document}


\title{Deep imaging of Eridanus~II and its lone star cluster\altaffilmark{*}}


\author{D. Crnojevi\'c\altaffilmark{1}, D. J. Sand\altaffilmark{1},
  D. Zaritsky\altaffilmark{2}, K. Spekkens\altaffilmark{3},
B. Willman\altaffilmark{2,4,5}, J. R. Hargis\altaffilmark{6}}





\begin{abstract}
We present deep imaging of the most distant
dwarf discovered by the Dark Energy Survey, Eridanus~II (Eri~II). Our
Magellan/Megacam stellar photometry reaches $\sim$$3$~mag
deeper than previous work, and allows us to confirm the 
presence of a stellar cluster whose position is consistent with 
Eri~II's center. This makes Eri~II, at $M_V=-7.1$, the least
luminous galaxy known to host a (possibly central) cluster.
The cluster is partially resolved, and at $M_V=-3.5$ it accounts for 
$\sim$$4\%$ of Eri~II's luminosity.
We derive updated structural parameters for Eri~II,
which has a half-light radius of $\sim$$280$~pc and is
 elongated ($\epsilon$$\sim$$0.48$), at a measured distance 
of $D$$\sim$$370$~kpc.
The color-magnitude diagram displays a blue, extended
horizontal branch, as well as a less populated red horizontal
branch.
A central concentration of stars brighter than the old main sequence
turnoff hints at a possible intermediate-age ($\sim$$3$~Gyr) 
population; alternatively, these sources could be blue
straggler stars. A deep Green Bank Telescope observation of Eri~II 
reveals no associated atomic gas.
\end{abstract}


\altaffiltext{*}{This paper includes data gathered with the 6.5 meter Magellan Telescopes at Las Campanas Observatory, Chile.}
\altaffiltext{1}{Department of Physics, Texas Tech University, Box 41051, Lubbock, TX 79409-1051, USA; \email{denija.crnojevic@ttu.edu}}
\altaffiltext{2}{Steward Observatory, University of Arizona, 933 North Cherry Avenue, Tucson, AZ 85721, USA}
\altaffiltext{3}{Department of Physics, Royal Military College of Canada, Box 17000, Station Forces, Kingston, ON K7L 7B4, Canada}
\altaffiltext{4}{LSST, University of Arizona, 933 North Cherry Avenue, Tucson, AZ 85721, USA}
\altaffiltext{5}{Departments of Physics and Astronomy, Haverford College, 370 Lancaster Avenue, Haverford, PA 19041, USA}
\altaffiltext{6}{Space Telescope Science Institute, 3700 San Martin Drive, Baltimore, MD 21218, USA}


\section{Introduction}

The last several years have seen another burst of discovery of Milky Way 
(MW) satellites from ATLAS \citep{belokurov14, torrealba16}, the Panoramic 
Survey Telescope \& Rapid Response System \citep[Pan-STARRS;][]{laevens15a, laevens15b},
the Dark Energy Survey \citep[DES;][]{bechtol15, koposov15, drlica15} 
and the Dark Energy Camera more generally \citep{kim15a, kim15b, Martin15}.  

The most distant of the new satellites is Eridanus~II (Eri~II), at $D$$\sim$350~kpc 
and $M_V$$\sim$$-7$~mag, which was discovered simultaneously by two groups 
\citep{bechtol15,koposov15}. The distance to Eri~II roughly corresponds 
to the virial radius of the MW \citep[$\sim$300 kpc; e.g.][]{klypin02}:
this places Eri~II near the transition radius
between those dwarf spheroidals (with $D\lesssim$300~kpc),
which appear to have lost their primordial gas,
and more distant dwarfs which generally retain HI gas reservoirs 
\citep[e.g.][]{spekkens14}. The properties of Eri~II suggest that it
may be similar to Leo~T, another recently discovered dwarf just beyond the virial radius of 
the MW \citep[$D$$=$410~kpc, $M_V\sim$$-8$;][]{jdejong08}, which 
displays a significant HI gas reservoir \citep{ryan08} and has undergone 
multiple epochs of star formation \citep{jdejong08, weisz12}.  Additionally, 
\citet{koposov15} tentatively suggested that Eri~II may have a globular cluster, 
which would make it the smallest known galaxy with its own star cluster.

Motivated by the hints that set Eri~II apart from the bulk of the new 
MW satellites, we obtained deep photometric and HI observations to further 
study its physical properties.  In Section~\ref{sec:reduction}, we describe the optical 
Magellan/Megacam observations, 
as well as our stellar photometry. In Section~\ref{sec:analysis}
we present the color-magnitude diagram (CMD) of Eri~II and its stellar
populations, we derive updated values for its distance, structural
parameters, luminosity and HI gas content, and we describe the 
properties of its stellar cluster. We discuss and 
conclude in Section~\ref{sec:discussion}.


\begin{figure*}
 \centering
\includegraphics[width=18cm]{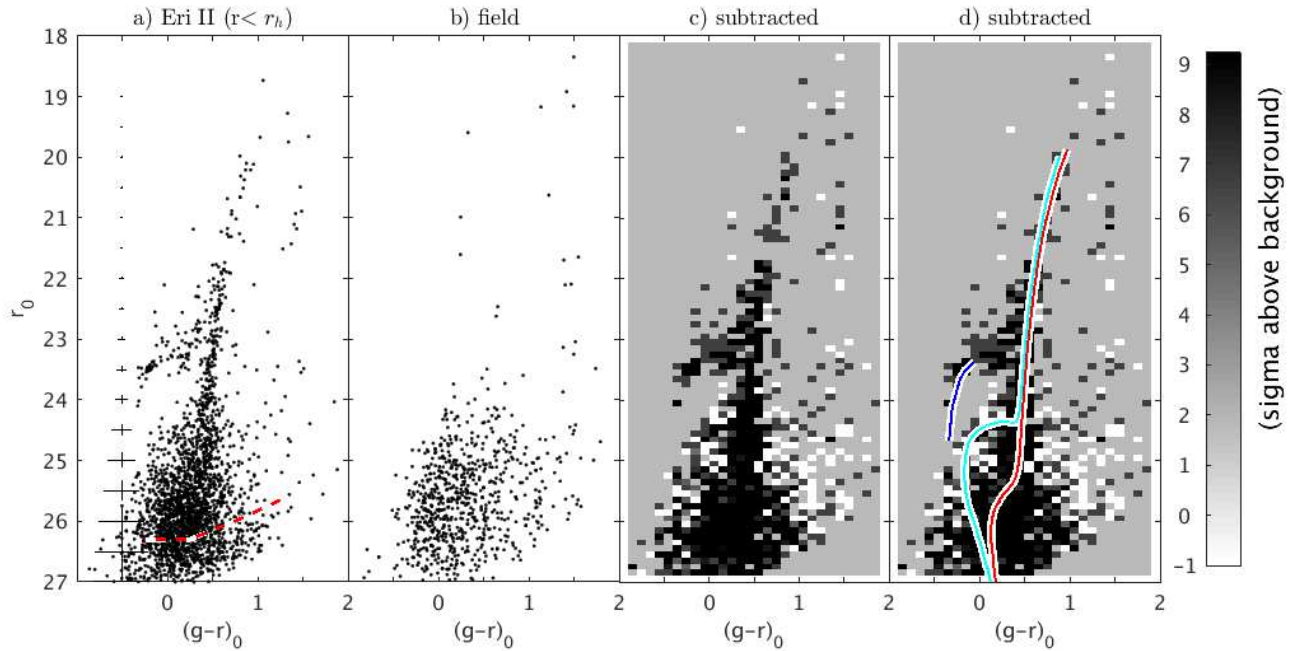}
\caption{\emph{Panel a)}. CMD
of Eri~II sources within its half-light radius. We show photometric 
errors and $50\%$ completeness levels (red dashed lines) as derived
from artificial star tests.
\emph{Panel b)}. Background field CMD 
rescaled to the area in panel a).
\emph{Panel c)}. Hess density diagram of the Eri~II CMD after 
background subtraction.
\emph{Panel d)}. Same as panel c), with
a 10~Gyr, [Fe/H]$=-2.5$ isochrone (red line),
a 3~Gyr, [Fe/H]$=-2.0$ isochrone (cyan line; \citealt{dotter08}), and the 
HB fiducial of M92 (blue solid line) overplotted, all shifted to the distance 
of Eri~II.
}
\label{cmd_glob}
\end{figure*}

\section{Data reduction and photometry}  \label{sec:reduction}

Eri~II was observed on 2015 October 12 (UT) with the 
Megacam imager \citep{mcleod15} on the Magellan Clay telescope. 
Megacam has a $\sim$$24'\times24'$ field-of-view and a binned
pixel scale of 0\farcs16. The seeing was 
$\sim$0\farcs9 and $\sim$0\farcs6 for the $g$- and $r$-band, respectively.
We took 14 300~sec exposures in $g$-band and 
6 300~sec exposures in $r$-band, which were dithered 
to fill in the gaps between the CCDs.
The data were reduced using the Megacam pipeline at the 
Smithsonian Astrophysical Observatory Telescope Data Center.

We perform point spread function fitting photometry on each of
the stacked final images, using the suite of programs DAOPHOT and 
ALLFRAME \citep{stetson94}, and following the procedure described in
\citet{crnojevic16}.
To calibrate our data, we used the AAVSO Photometric 
All-Sky Survey \citep[APASS;][]{henden12} $g$- and $r$-band 
photometry and cross-matched it with short (60~sec) Eri~II exposures, 
propagating the derived zeropoints and color terms 
to the longer science exposures.  The calibrated catalogues were corrected 
for Galactic extinction \citep{schlegel98, schlafly11} star by star
(the average extinction across the image is $\sim0.01$~mag), and 
we present dereddened $g_0$ and $r_0$ magnitudes throughout this work. 

We performed a series of artificial star tests to determine our 
photometric errors and completeness as a function of magnitude and color.  
Over many experiments, we injected $\sim$10 times the number of artificial 
stars into our data as were detected in the unaltered images. The artificial stars
were homogeneously distributed across the field-of-view, with a 
color-magnitude distribution similar to that of the observed sources, 
but extending $\sim$2 magnitudes fainter.   
The photometry of the images with artificial star injections was 
derived exactly in the same way as for the real data. 
The color-averaged $50\%$ completeness limits are $r_0$$\sim$26 
and $g_0$$\sim$26.6~mag. In Fig.~\ref{cmd_glob}, we present the  
CMD of Eri~II, along with our derived photometric errors and $50\%$ completeness
limits.


\section{Analysis} \label{sec:analysis}

The CMD for sources within Eri~II's half-light radius ($r_h$) is presented 
in Fig.~\ref{cmd_glob}, along with a background CMD 
derived from two regions of identical size located away from Eri~II. 
Our CMD reaches $\sim$$3$~mag deeper than the discovery data 
\citep{bechtol15,koposov15}.

\begin{figure}
 \centering
\includegraphics[width=6cm]{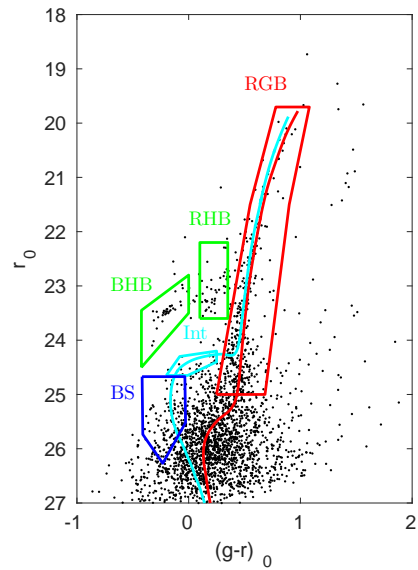}
\caption{CMD of Eri~II, showing the selection boxes used to select
different subpopulations. Isochrones are as in Fig.~\ref{cmd_glob}.
}
\label{boxes}
\end{figure}

\begin{figure*}
 \centering
\includegraphics[width=7cm]{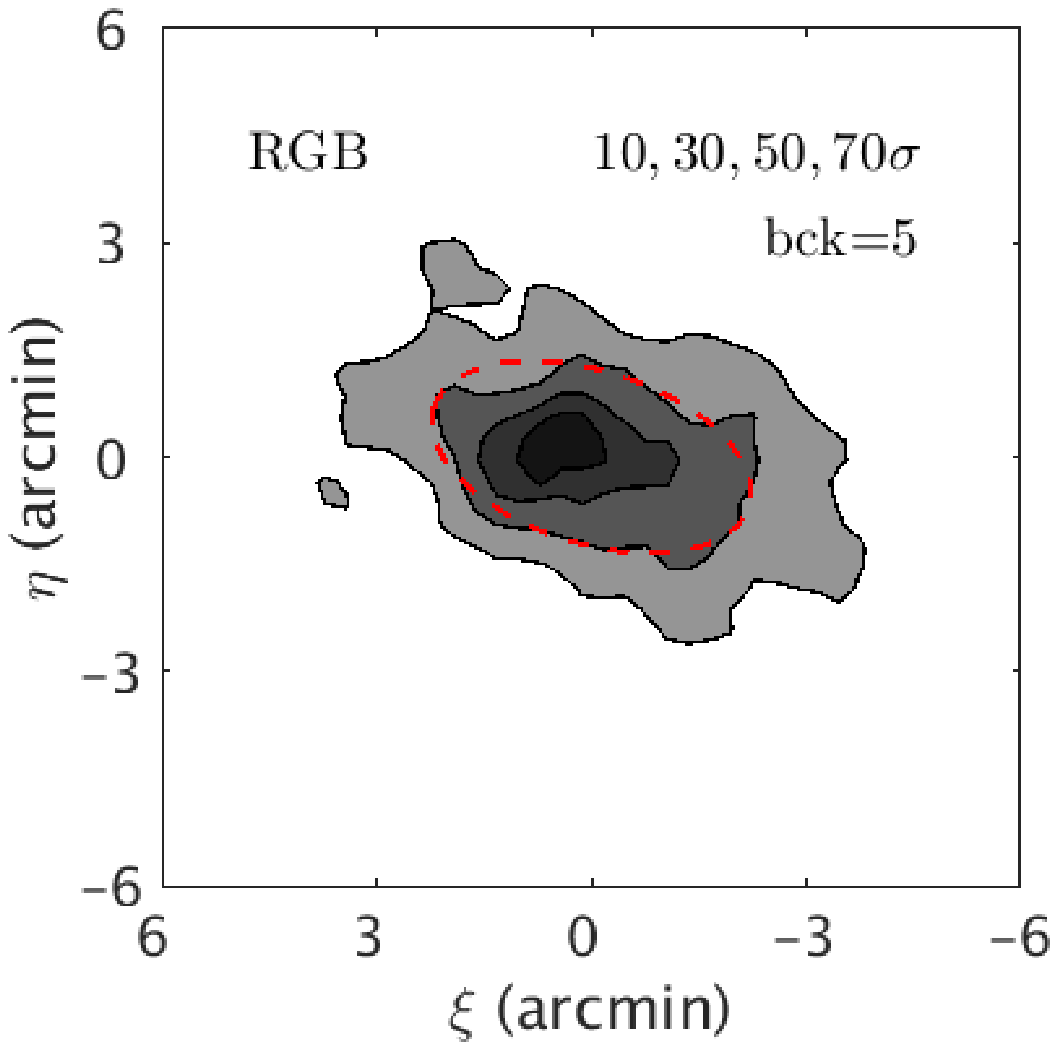}
\includegraphics[width=7cm]{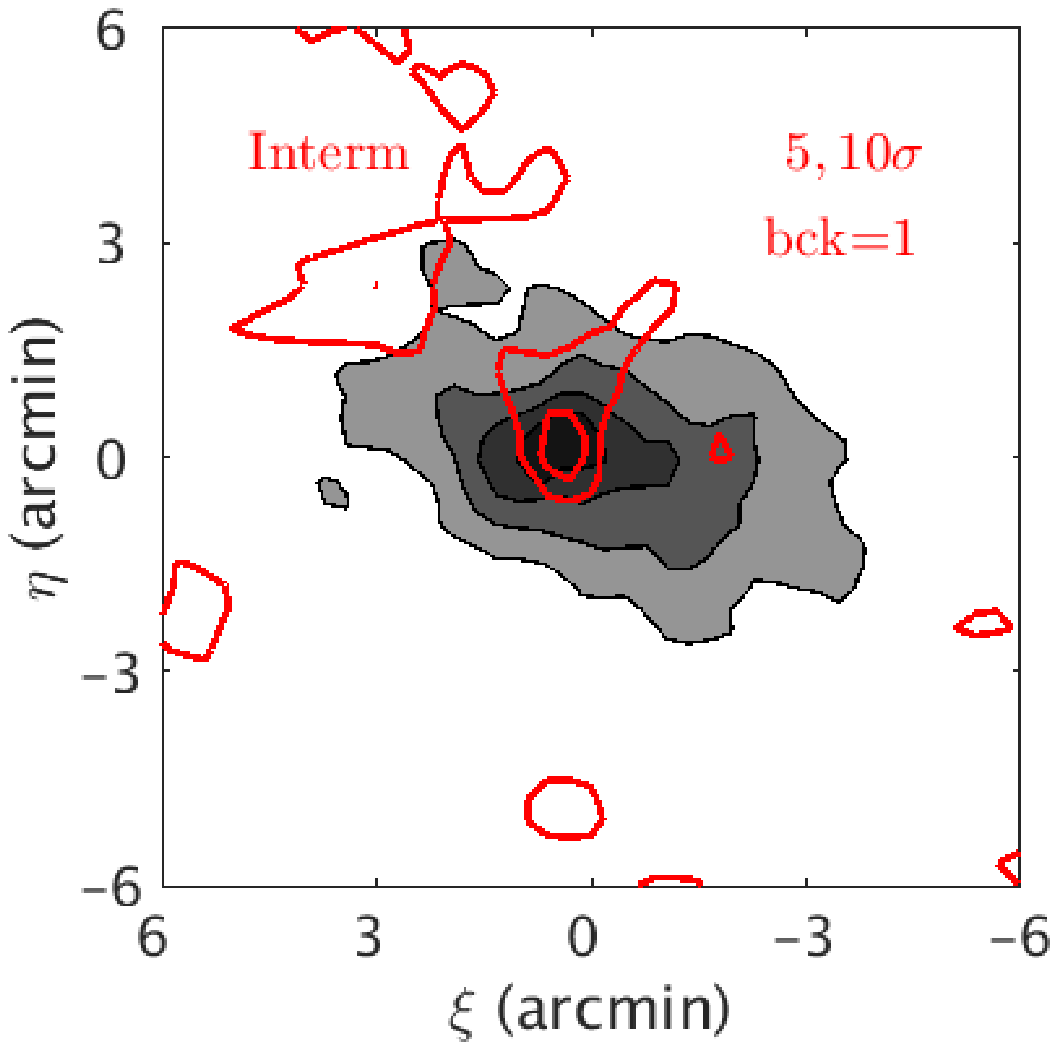}
\includegraphics[width=7cm]{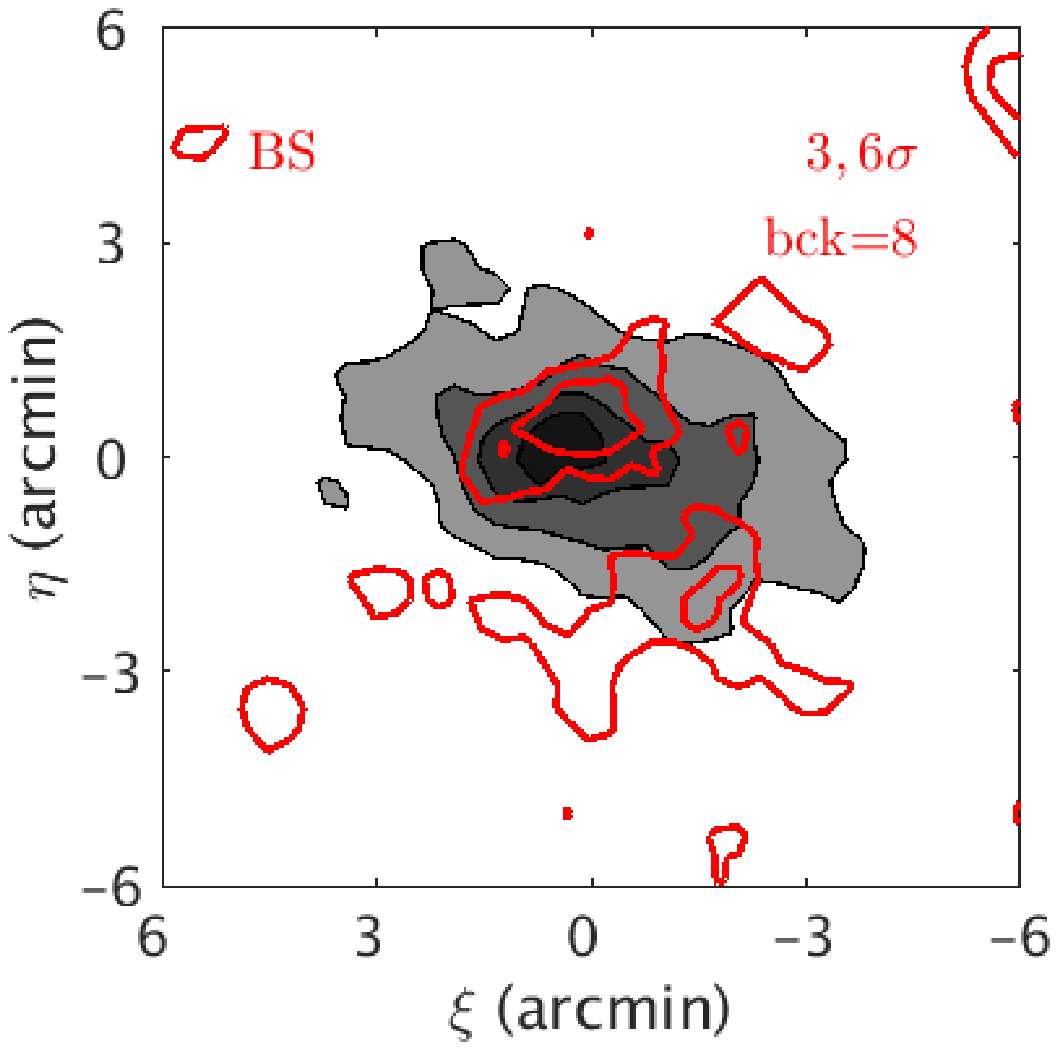}
\includegraphics[width=7cm]{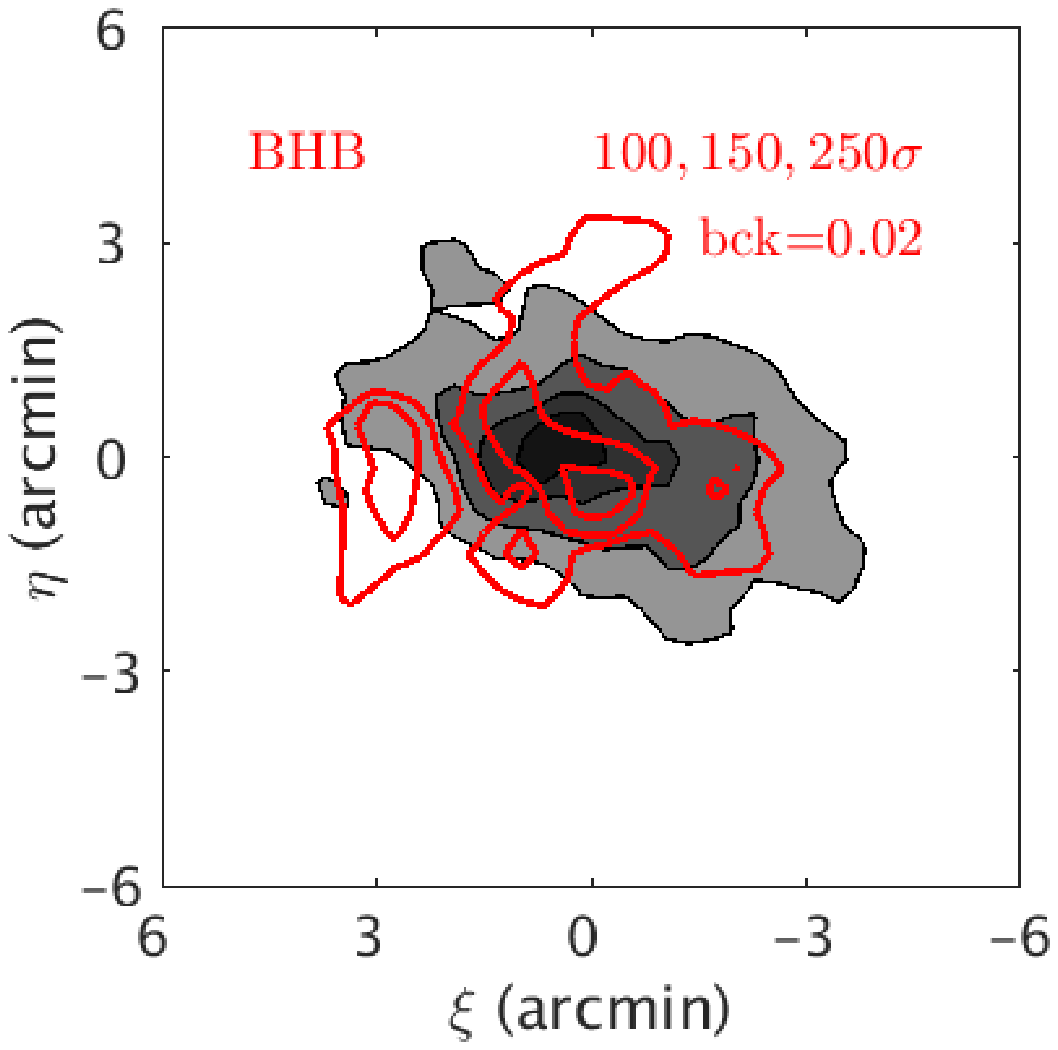}
\includegraphics[width=7cm]{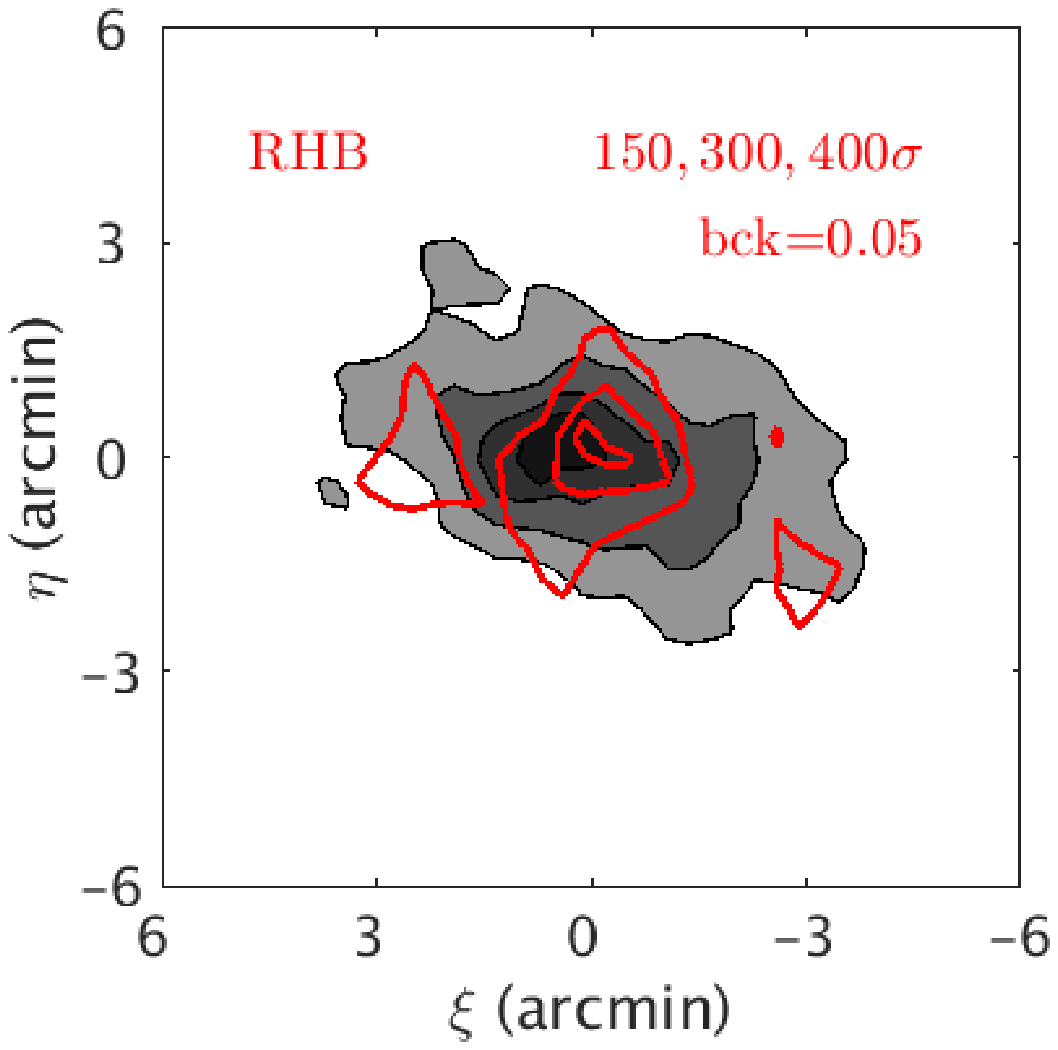}
\caption{Stellar maps of different CMD features. 
The red dashed ellipse in the \emph{top left panel}
indicates $r_h$. Isodensity contours are 
logarithmically spaced and indicate $\sigma$ above the field level,
with values reported in each subpanel; indicated is also the
field level in units of stars per arcmin$^2$.
The subpopulations are: RGB stars 
(thin contours and filled distribution in each subpanel), 
intermediate-age subgiant branch stars 
(red contours in \emph{top right panel}), candidate BS stars 
(red contours in \emph{central left panel}),
BHB stars (red contours in \emph{central right panel}), 
and RHB stars (red contours in \emph{bottom panel}).
}
\label{map}
\end{figure*}

\subsection{Stellar populations} \label{sec:cmds}

Eri~II is not simply a single old and metal poor stellar population, 
in contrast to many of the new, faint satellites of the MW \citep{brown14}.
The color and slope of the RGB and the blue extent of the
extended horizontal branch (HB; Fig.~\ref{cmd_glob}) are characteristic of old, 
metal-poor ($\gtrsim10$~Gyr, [Fe/H]$\lesssim-1.0$) populations such as those in 
globular clusters or ultra-faint dwarfs.
The HB is not purely blue, but has a red HB (RHB) portion as well:
while this can be a feature of old and metal-rich populations,
the RGB slope of Eri~II seems to exclude the presence
of significantly enriched stars. Alternatively, the RHB could arise
from a metal-poor population marginally younger ($\sim2$~Gyr) than the 
oldest one, or as a second parameter effect \citep{salariscassisi05}.
The objects between the BHB and the RHB at a 
range of magnitudes are candidate RR~Lyrae stars.

We consider the Hess diagram for the field-subtracted
CMD (Fig.~\ref{cmd_glob}): an excess of sources is present above the subgiant 
branch portion of the old isochrone, around $-0.20\lesssim (g-r)_0\lesssim0.15$ 
and $24.3\lesssim r_0\lesssim25.3$.
This feature is well fit by a $\sim$$3$~Gyr and [Fe/H]$=-2.0$ isochrone
(panel d of Fig.~\ref{cmd_glob}), or alternatively by a 
$\sim$$4$~Gyr population as metal-poor as the old RGB.
A significantly younger and more metal-rich population can be ruled out,
since it would produce a RGB redder than observed; similarly,
for older ages the turnoff would be too faint to match the observed one. 
We cannot exclude the possibility that these are,
at least in part, blue straggler (BS) stars arising from binary systems, which 
can mimic intermediate-age populations \citep[for a discussion about
the ``blue plume'' in faint MW satellites see][]{sand10,
santana13, weisz15}. The ratio of the number of candidate 
BS stars to the number of blue HB (BHB) stars (computed
following \citealt{deason15}) is $\sim$$1.7\pm0.4$, 
consistent with empirical values derived for dwarf galaxies with 
luminosities similar to Eri~II \citep[e.g.,][]{deason15}. This
supports the BS interpretation for these stars.

We select different CMD populations as shown in Fig.~\ref{boxes}, and
draw their spatial distribution (as density maps) in Fig.~\ref{map}. 
RGB stars present a regular and elongated shape, with no clear 
signs of disturbance. Intermediate-age stars
produce a very noisy stellar density map but have a significant
overdensity at the center of Eri~II. The map for candidate
BS stars is similarly noisy, but it still shows a possible central
concentration, which would favor the intermediate-age
population interpretation (BS binary systems are expected
to have a spatial distribution similar to that of the old population).
The BHB and RHB maps have a very high significance above
the field level; the BHB sample's highest overdensity is
slightly offset from Eri~II's center. 

\citet{koposov15} suggested that Eri~II possesses a young stellar 
population component ($\sim$$250$~Myr) based on the presence of a few stars
blueward of the RGB's bright end (at $r_0\sim$$20$ and 
$(g-r)_0\sim$$0.2$).
If these were truly young stars, we would detect sources along
the main sequence portion (around $r_0\gtrsim22.5$) of the young isochrone 
plotted in Fig.~16 of \citet{koposov15}, but we do not observe these
in our dataset.

\subsection{Distance} \label{sec:dist}

We rederive the distance to Eri~II from the luminosity of its BHB \citep[e.g.,][]{sand12}. 
We adopt the fiducial HB sequence for M92 from 
\citet[][]{bernard14b},
and convert the Pan-STARRS1 magnitudes into the SDSS system
following \citet{tonry12}.
While we adopt the nominal M92 distance value reported by \citet{bernard14b}
\citep[$(m-M)=14.65$, from][]{harris10}, other studies have
measured slightly different values, as high as $(m-M)=14.74$ 
\citep[see][]{dicecco10}, but they are
all within their respective $1\sigma$ errors.
We thus adopt an M92 distance uncertainty of $\sigma_{(m-M),M92}=0.1$,
which is the major source of error in our measurement.
We perform a least-squares fitting of the M92 HB fiducial to
 Eri~II's BHB, and we find a best-fit value of 
$(m-M)_0=22.8\pm0.1$, corresponding to a distance of $D=366\pm17$~Mpc.
This is consistent with the results of \citet{koposov15}.

\subsection{Structural properties and luminosity} \label{sec:prof}

We re-determine structural parameters for Eri~II 
with the same maximum likelihood technique utilized for most of the other 
Local Group (LG) dwarfs \citep[][]{martin08, sand12}. 
RGB stars were included in our structural analysis;
we did not explicitly include the HB or potential younger stellar 
population, although any true younger stellar population would 
inevitably have RGB stars that would also be included. We did not mask
the central star cluster for our structural analysis,
although a separate calculation with the star cluster masked out
led to structural parameters within 1-$\sigma$ of the original. The free 
parameters of our exponential profile plus constant background fit were:
central position, position angle, ellipticity, $r_h$ and 
background surface density (results are reported in Table~\ref{tab1}).
Uncertainties were determined by bootstrap resampling the data 1000 times, 
recalculating the structural parameters for each resample.

Eri~II's surface brightness profile is well fit by an
exponential with $r_h=277\pm14$~pc, and its
ellipticity is high ($\epsilon=0.48\pm0.04$), similar to other
faint MW satellites \citep[][]{sand12}. The central surface
brightness is derived from the total luminosity (see below),
$r_h$ and the ellipticity, and it is $\mu_{V,0}=27.2\pm0.3$~mag/arcsec$^2$.
While most of the derived parameters are in line with the discovery results, our $r_h$
is higher by a factor of $\sim$$1.5$:  deep follow-up photometric studies 
often lead to increased $r_h$ 
values with respect to shallower datasets \citep[e.g.][]{crnojevic14a}.

Given the likely presence of two subpopulations, we derive the 
absolute magnitude of Eri~II as follows. We derive the 
completeness-corrected and field-subtracted luminosity for each 
subpopulation by summing the flux of stars in a magnitude range where 
the corresponding isochrones
do not overlap within the observational errors 
($24.7<r_0<25.5$ for the intermediate-age population and $24.4<r_0<25.5$
for the older population). The luminosity is computed within
$r_h$ and then multiplied by two.
We extrapolate the total luminosity given by each component 
starting from the fraction of light contributed in the selected magnitude 
ranges, as estimated from theoretical luminosity functions 
from \citet{dotter08}. These are constructed with a Salpeter initial mass 
function, one for a 10~Gyr, [Fe/H]$=-2.5$ population and the other for a 
3~Gyr, [Fe/H]$=-2.0$ population. 
The luminosities associated with the subpopulations are
$L_{V,old}\sim$$5.6\pm1.5\times10^4 L_\odot$ and 
$L_{V,int}\sim$$3.5\pm3.0\times10^3 L_\odot$:
the intermediate-age population contributes 
$\sim$$6\%$ of Eri~II's light, and only a few percent of its stellar mass.
The total absolute magnitude for Eri~II is
$M_{V}=-7.1\pm0.3$ (the $V$-band magnitude was obtained
using the \citealt{jester05} transformations), which is 
between the values reported by \citet{bechtol15} and 
\citet{koposov15}. In case the stars
brighter than the old subgiant branch had a BS component, the absolute 
magnitude of Eri~II would not significantly change, given their small number.

\begin{figure*}
 \centering
\includegraphics[width=6cm]{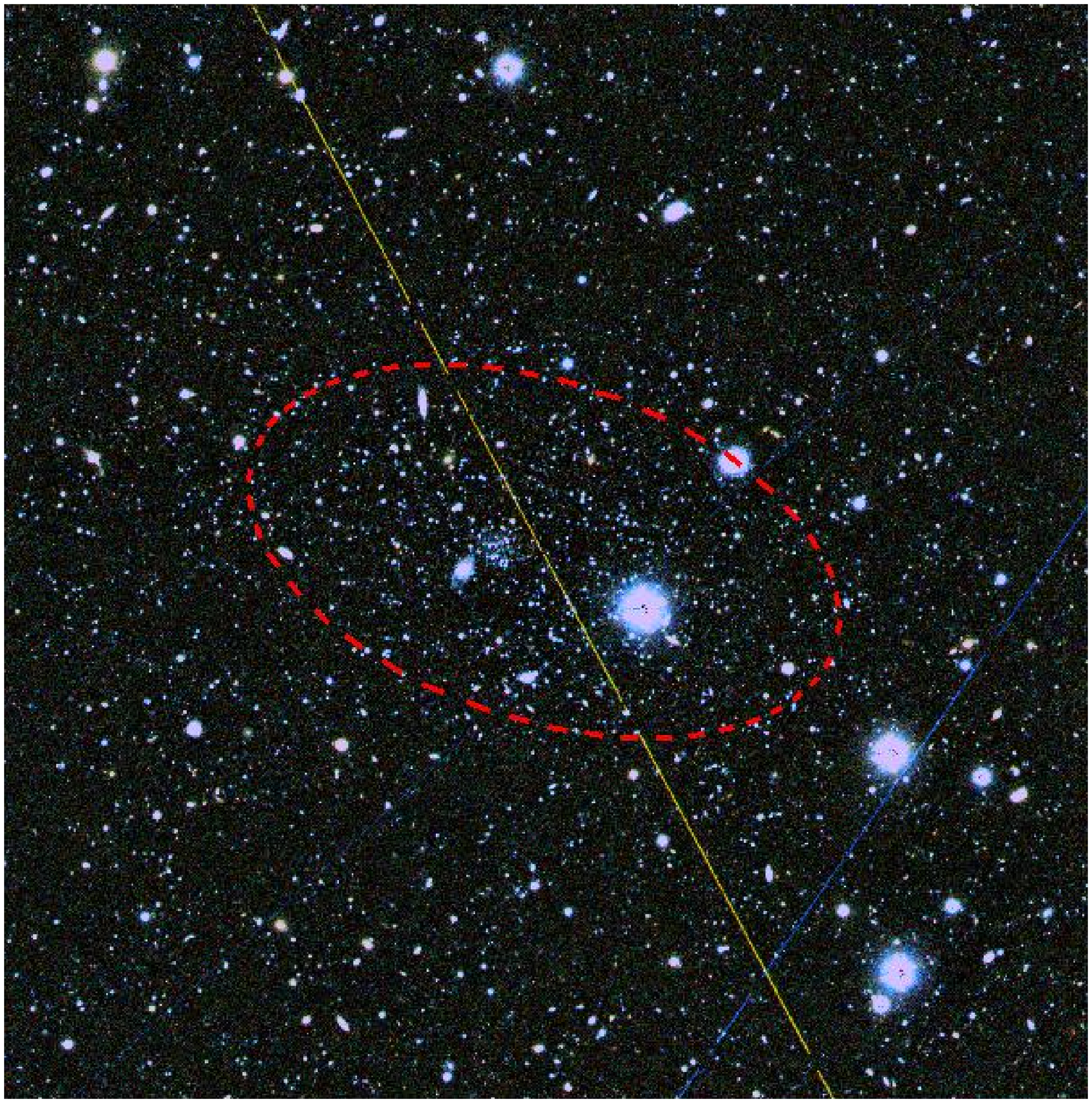}
{\includegraphics[width=6.cm]{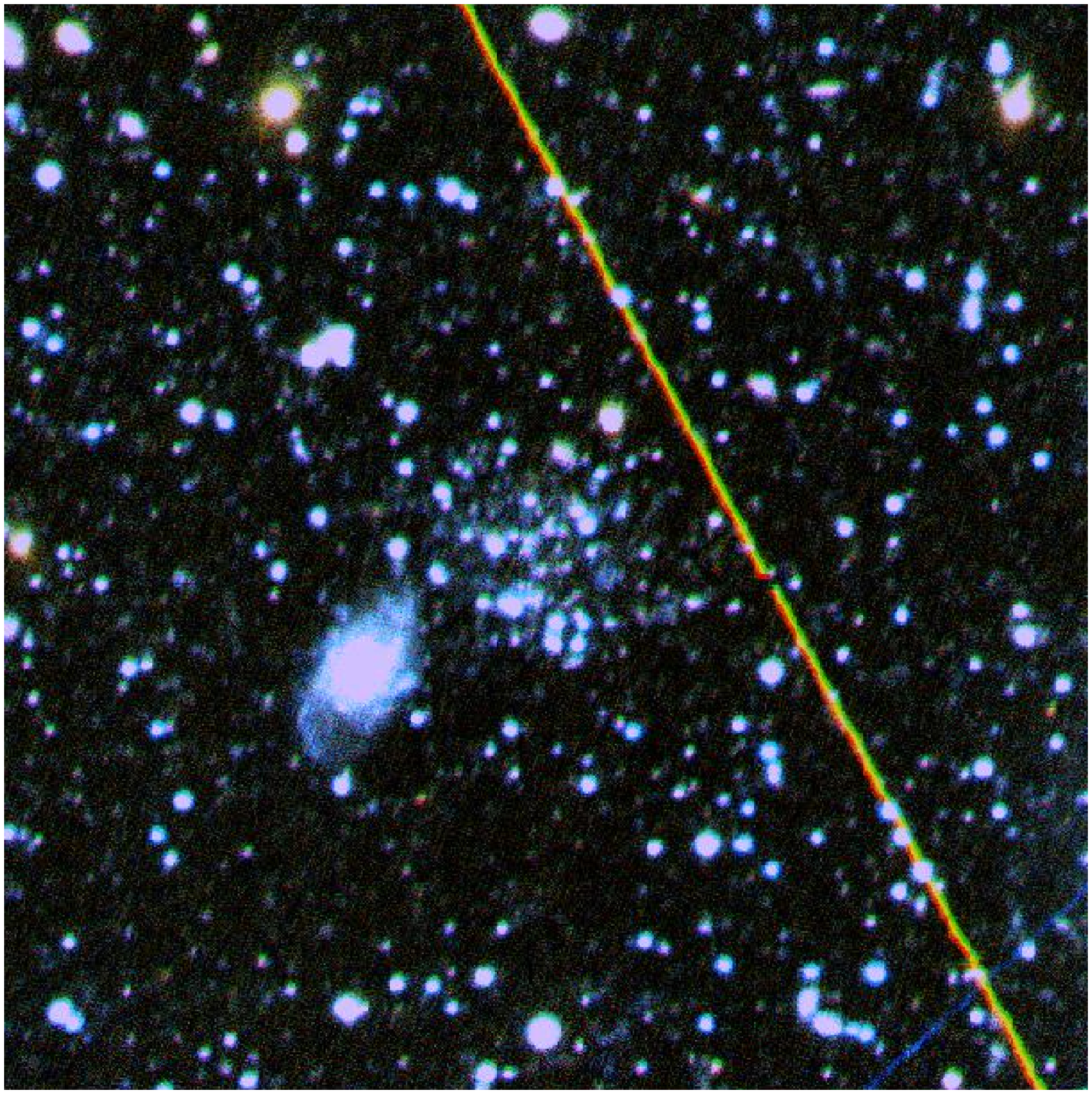}}
\includegraphics[width=6cm]{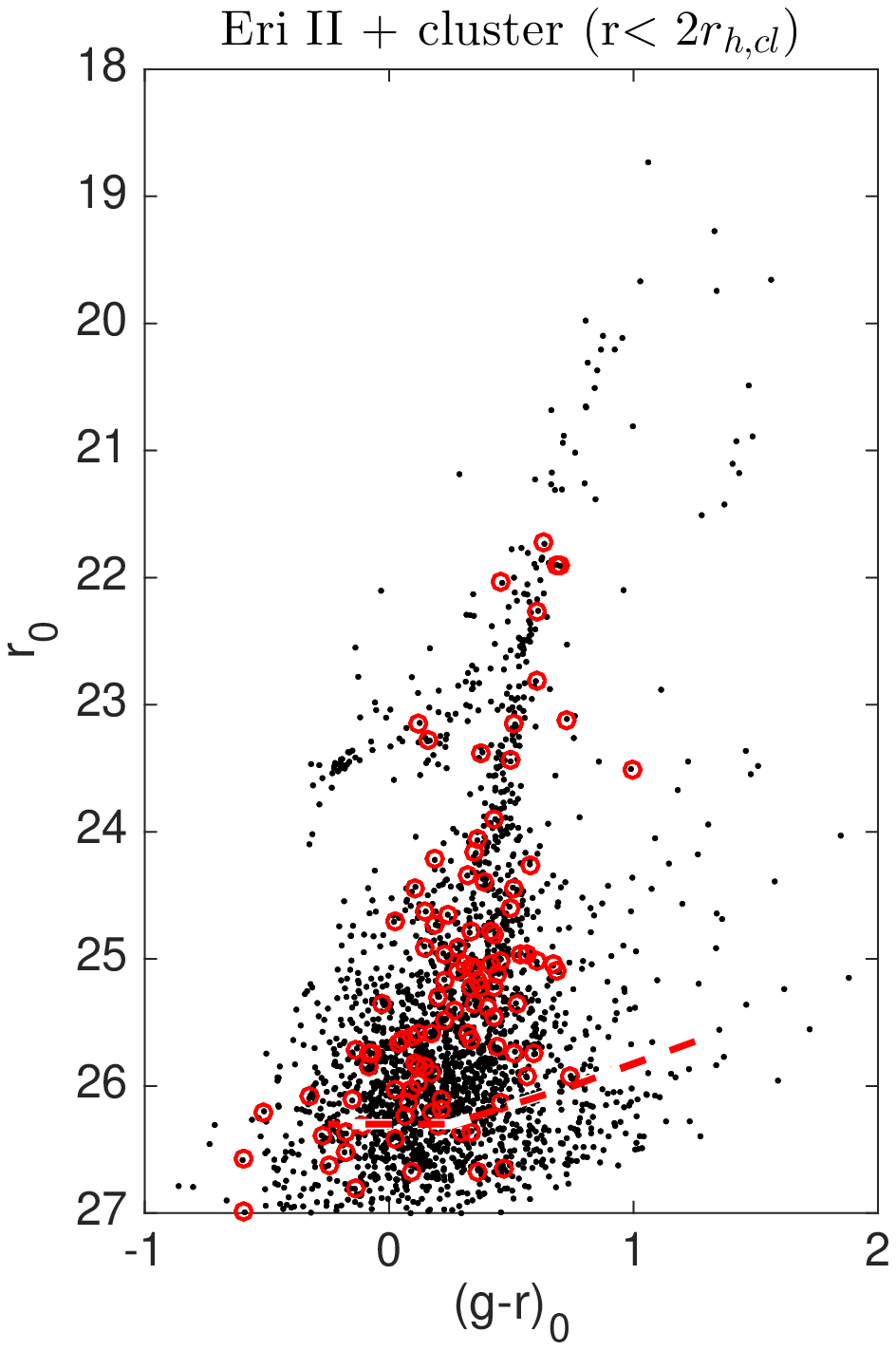}
\caption{RGB-color images of Eri~II. 
A satellite trail is apparent in both panels.
\emph{Upper left panel}. $10'\times10'$ field centered on Eri~II;
a red dashed ellipse indicates $r_h$.
\emph{Upper right panel}. $2'\times2'$ zoom-in on Eri~II's
cluster, which is partially resolved into stars. 
\emph{Lower panel}. Dereddened CMD for Eri~II within $r_h$;
in red we overlay stars from the cluster within twice its 
computed $r_h$.
}
\label{images}
\end{figure*}

\subsection{HI gas content} \label{sec:gas}

To constrain the HI content of Eri~II, we obtained position-switched HI
observations (AGBT-16A-186; PI: Spekkens) on the 
Robert C. Byrd Green Bank Telescope (GBT)
on 2016 Feb 23. In the range $-500 \leq V_{LSRK} \leq -50$~$\kms$
and $50 \leq V_{LSRK} \leq 500$~$\kms$ -- encompassing possible HI
recessional velocities of MW satellites that do not overlap
with the MW HI emission itself along the Eri~II line of sight --
the GBT spectrum has an rms noise $\sigma = 1.2 \,$mJy at a spectral
resolution of $15\,\kms$. We do not find any HI emission in this
velocity range within the $\mathrm{FWHM} = 9.1\arcmin =
960\,\mathrm{pc}$ GBT beam at this frequency. Combining this
non-detection with the distance and luminosity values derived above,
a putative HI counterpart has a 5$\sigma$, $15\,\kms$ HI mass upper
limit of $\Mlim = 2.8 \times 10^3\,\Msun$ out to $\sim$$1.5 r_h$ ,
and therefore $\gasfrac < 0.036 \, \MLsun$; this constraint is a
factor of $\sim$7 more stringent than that derived by
\citet{westmeier15} using HIPASS data. Eri~II is therefore extremely gas-poor,
similar to other dwarf spheroidal galaxies in the Local Volume 
\citep[][]{grcevich09, spekkens14}.

\subsection{The central star cluster} \label{sec:cluster}

A dense collection of stars near Eri~II's center stand out from the overall 
diffuse nature of the dwarf \citep[Fig.~\ref{images}; 
see also][]{koposov15}.
This likely star cluster makes Eri~II, at 
$M_{V}\sim$$-7.1$~mag, by far the least luminous dwarf galaxy to host such an object.   
The position of the cluster is nearly at the calculated center of 
Eri~II ($\sim$$45$~pc off-center in projection), and within 
its uncertainties (see Tab.~\ref{tab1}).
Small offsets between centrally-located clusters and the exact 
center of their parent dwarf galaxy are commonly observed 
\citep[e.g.,][]{georgiev09}, so we plausibly consider
this as a candidate nuclear star cluster, more specifically the only
one known within such a faint galaxy.

In Fig.~\ref{images}, we overlay on Eri~II's CMD
the stars contained within twice the $r_h$ of its cluster
(see below). The center of the cluster was computed iteratively as the average of 
the stellar positions within circles of decreasing radius.
Since the cluster is only partially resolved, its CMD is poorly 
populated, but still consistent with Eri~II's CMD.
To test this, we randomly draw 1000 sub-CMDs with the same
number of sources as in the cluster from Eri~II's CMD: the results
resemble the cluster's CMD, and $\sim15\%$ of the realizations
also lack a BHB.
We compute the cluster's properties via integrated
photometry, assuming a circular radius (the cluster is visually
round). The surface brightness
profile for the cluster is derived after masking bright stars and
background galaxies, and is then fit with a Sersic profile
(best-fit values are reported in Tab.~\ref{tab1}).
The absolute magnitude is derived by integrating the best-fit
Sersic profile and is $M_V=-3.5\pm0.6$ at
the distance of Eri~II, which contributes to $\sim$$4\%$ of its host's
luminosity. We discuss the properties of the cluster in the next section.

\begin{deluxetable}{lcc}
\tablecolumns{3}
  \tablecaption{Properties of Eri~II and its cluster.}

\tablehead{
\colhead{Parameter}  & \colhead{Eri~II} & \colhead{Cluster} \\
}\\

\startdata
RA (h:m:s) & 03:44:20.1$\pm10.5$'' & 03:44:22.2$\pm1$'' \\
Dec (d:m:s) & $-43$:32:01.7$\pm5.3$'' & $-43$:31:59.2$\pm2$'' \\
$(m-M)_0$ (mag) & $22.8\pm0.1$ & -- \\
D (kpc) & $366\pm17$ & -- \\
$\epsilon$ & $0.48\pm0.04$ & -- \\ 
PA (N to E; $^o$) & $72.6\pm3.3$ & -- \\
$r_{h}$ (arcmin) & $2.31\pm0.12$ & $0.11\pm0.01$  \\
$r_{h}$ (pc) & $277\pm14$ & $13\pm1$  \\
$n$ (Sersic index)  & $1\tablenotemark{a}$ & $0.19\pm0.05$ \\
$\mu_{V,0}$ (mag/arcsec$^2$) & $27.2\pm0.3$ & $25.7\pm0.2$  \\
$M_V$ (mag) &$-7.1\pm0.3$ & $-3.5\pm0.6$ \\
$<$$(g-r)_0$$>$ (mag) &$0.5\pm0.3$ & $0.4\pm0.2$ \\
$\gasfrac$ ($\MLsun$) & $< 0.036$ & -- \\
\enddata
\tablenotetext{a}{An exponential profile was assumed for Eri~II.}
\label{tab1}

\end{deluxetable}


\section{Discussion and conclusions}  \label{sec:discussion}

We have presented deep Magellan/Megacam photometric 
follow-up observations of Eri~II, and derived improved distance, structural 
properties, and luminosity measurements.
At a distance of $370$~kpc, Eri~II is just beyond 
the estimated MW virial radius, where most dwarfs show signs
of neutral gas/recent star formation. However, Eri~II 
is not currently forming stars (based on the lack of 
a blue main sequence), unlike the two gas-rich dwarfs
found at a similar Galactocentric distance, Leo~T 
\citep[$M_V\sim$$-8.0$; e.g.,][]{weisz12} and
Phoenix \citep[$M_V\sim$$-10.0$; e.g.,][]{battaglia12}.
Our follow-up GBT observations reveal that Eri~II is extremely gas-poor. 
This suggests that it is not on first infall and may have already interacted with
the MW. The stellar distribution shows no signs of disruption down to
the current photometric depth, disfavoring tidal interactions. Rather,
the stellar and gas properties of Eri~II are consistent with a ram
pressure stripping scenario in which its gas reservoir was swept away
by the MW's coronal gas during a previous pericentric passage 
\citep[e.g.,][]{gatto13}.

At $M_V=-7.1$, Eri~II is the faintest dwarf galaxy 
known to host a stellar cluster, consistent with being at the center of its host. 
Compared to the LG's globular clusters, Eri~II's cluster would be one of
the faintest objects at its $r_h$ value, and its absolute magnitude and $r_h$ 
place it in a similar parameter space to the least luminous and 
most compact MW ultra-faint dwarfs.
With $r_h=13$~pc, it can be regarded as an ``extended''
cluster: these are only observed at large galactocentric distances
in M31, or in dwarf galaxies like NGC~6822 and Scl-dE1
\citep[see][]{dacosta09, mackey10}. Such objects are believed to 
survive only within weak tidal fields \citep[e.g.,][]{hurley10}.

Is Eri~II's cluster a survivor?
We compute its half-mass relaxation time based on our derived
luminosity, assuming a mass-to-light ratio of $\sim$$1.5$, and 
obtain $\sim$$2$~Gyr \citep[e.g.,][]{dacosta09}. Since the 
timescale for evaporation due to two-body relaxation is on the order 
of $\gtrsim10$ times this quantity, internal processes must not be relevant 
for our object, and the weak tidal field of Eri~II likely does not
contribute to a possible tidal truncation. A disruption could happen in presence
of strong external tidal forces, however the lack of tidal features
in Eri~II and its large Galactocentric distance make
this an unrealistic possibility.

Nuclear star clusters in dwarf galaxies likely form via inspiral
due to dynamical friction, rather than being born in situ 
\citep[see, e.g.,][]{denbrok14}, even though
this interpretation is far from definitive.
Around the MW and M31
only a handful of satellites host clusters, the faintest being Fornax
($M_V\sim$$-13.5$) and And~I ($M_V\sim$$-11.7$; \citealt{mcconnachie12}),
but none of these seem to be a nuclear star cluster. 
In Fornax, the presence of a globular cluster system and the lack
of a nuclear cluster suggest that the dynamical friction timescale
is prolonged by the existence of a cored, rather than cuspy, dark matter 
profile \citep{goerdt06}.
Beyond the LG, a few dwarfs as faint as $M_V\sim$$-10$ host 
candidate globular clusters \citep{dacosta09, georgiev09};
however, a deep study of the Fornax galaxy cluster found 
no galaxies with $M_i>$$-10$ (out of a sample of $\sim60$)
with nuclear star clusters \citep{munoz15}.
Clearly, the cluster in Eri~II is a rare event: this will be
key to investigate the formation of clusters
in such faint galaxies, and the modeling of its dynamical friction timescales
may help constrain Eri~II's dark matter profile.

Why, finally, are there no other dwarfs fainter than $M_V\sim$$-10$
with confirmed clusters? \citet{zaritsky16} suggest 
a scenario where at least a fraction of the outer halo Galactic clusters may
in fact reside in extremely low stellar density subhalos (like
Eri~II), which have so far gone undetected.
This uncharted territory certainly deserves further exploration and
detailed modeling.


\section*{Acknowledgements}

We thank the anonymous referee for their thoughtful suggestions,
and Tom Maccarone for useful discussions.
DJS acknowledges financial support from NSF grant AST-1412504;
DZ acknowledges support from NSF AST-1311326;
BW acknowledges support from NSF AST-1151462.
This paper uses data products produced by the OIR Telescope
Data Center, supported by the Smithsonian Astrophysical
Observatory.
The National Radio Astronomy operates the GBT,
and is a facility of the NSF operated under
cooperative agreement by Associated Universities, Inc.


 \bibliographystyle{apj}

\clearpage


\end{document}